\documentclass[manuscript]{aastex}
\usepackage{epsfig}
\usepackage{rotating}
\usepackage{lscape}

\shorttitle{Radio and X-ray properties of submm galaxies in the A2125 field}
\shortauthors{Wagg et al.}
\slugcomment{Accepted for publication in ApJ}

\def\ga{\mathrel{\raise0.35ex\hbox{$\scriptstyle >$}\kern-0.6em
\lower0.40ex\hbox{{$\scriptstyle \sim$}}}}
\def\la{\mathrel{\raise0.35ex\hbox{$\scriptstyle <$}\kern-0.6em
\lower0.40ex\hbox{{$\scriptstyle \sim$}}}}

\begin{document}
\title{Radio and X-ray properties of submillimeter galaxies in the A2125 field}

\author{
J. Wagg,$^{2,5}$  F. Owen,$^{2}$ F. Bertoldi,$^{3}$
M. Sawitzki$^{3}$ C.~L. Carilli,$^{2}$
 K.M.~Menten,$^{4}$ and H.~Voss$^{3,4}$}

\affil{$^{2}$National Radio Astronomy Observatory, PO Box O, Socorro, NM, USA 87801}

\affil{$^{3}$Radioastronomisches Institut der Universit{\"a}t Bonn, Auf dem H{\"u}gel 71, 53121 Bonn, Germany}

\affil{$^{4}$Max-Planck Institut f{\"u}r Radioastronomie, Auf dem H{\"u}gel 69, D-53121 Bonn, Germany}

\affil{$^{5}$Max-Planck/NRAO Fellow}

\begin{abstract}
We present the radio and X-ray properties of 1.2~mm MAMBO source
 candidates in a 1600~sq.~arcmin field centered on the Abell 2125 galaxy 
cluster at $z = 0.247$. 
 The brightest, non-synchrotron mm source candidate
 in the field has a photometric redshift, $z = 3.93^{+1.11}_{-0.80}$,
 and is not detected in a 31~ks \textit{Chandra} X-ray exposure. 
These findings are consistent with
 this object being an extremely dusty and luminous starburst galaxy at
 high-redshift, possibly the most luminous yet identified in any
 blank-field mm survey. 
The deep 1.4~GHz VLA imaging identifies counterparts for
 83\% of the 29 mm source candidates identified at $\ge$4-$\sigma$
 ($S_{1.2mm} = 2.7 - 52.1$~mJy), 
implying that the majority of these objects are likely to lie at $z \la 3.5$. 
The median mm-to-radio wavelength photometric redshift of this radio-detected 
sample is $z \sim 2.2$ (first and third quartiles of 1.7 and 3.0), 
consistent with the median redshift derived from optical spectroscopic
 surveys of the radio-detected subsample of bright
  submm galaxies ($S_{\rm 850~\mu m} > 5$~mJy). 
 Three mm-selected quasars are
 confirmed to be X-ray luminous in the high resolution \textit{Chandra} imaging, while
 another mm source candidate with potential multiple radio counterparts is also
 detected in the X-ray regime. Both of these radio
  counterparts are positionally consistent with the mm source
 candidate. One counterpart is associated with an elliptical galaxy at $z = 0.2425$,
  but we believe that a second counterpart associated with a fainter
 optical source likely gives rise to the mm emission at $z \sim 1$. 
\end{abstract}


\section{Introduction}

The discovery of massive, dust-obscured starburst galaxies in blank-field 
submm/mm-wavelength surveys has opened a new window on galaxy and structure formation 
 in the early Universe 
 (Smail, Ivison \& Blain 1997; Hughes et al.\ 1998; Barger et al.\ 1998; Bertoldi et al.\ 2000).
 The bulk of this population exists at redshift, $z$ $> 2$ (Chapman et al.\ 2003, 2005; 
 Aretxaga et al.\ 2003, 2007). These (sub)mm galaxies (hereafter SMGs) exhibit 
star-formation rates in excess of 1000~M$_{\odot}$/yr, sufficient to build up the stellar mass of 
 a giant elliptical galaxy in approximately 1~Gyr. This conclusion relies on the assumption that
 most of the
 far-infrared (FIR) luminosity in these objects is powered by star-formation rather than by an 
 active galactic nucleus (AGN). X-ray studies of these SMGs confirm that while $\sim$28--50\% do 
harbour  an AGN,
 such AGN activity may only contribute to a small fraction of the enormous FIR luminosities 
(e.g. Alexander et al.\ 2005). Deep radio imaging of the
  1.4~GHz emission in a subset of SMGs shows that this emission is extended
  over scales of a few kpc, also consistent with being associated with 
  star-formation activity (e.g. Biggs \& Ivison 2008).
 Further analysis of the X-ray and mid-infrared properties of SMGs are needed to reach stronger 
statistical conclusions 
regarding their AGN and star-forming nature, and such progress relies on the crucial first step of 
 obtaining accurate positional information allowing unambiguous identification of a source at different 
wavelengths.

The coarse angular resolution of submm/mm detectors on single-dish telescopes (typically 
$\sim$10--20$''$) means that it is impossible to immediately identify the optical/infrared counterpart 
responsible for the significant FIR luminosity of a SMG. Ideally one would like to use submm/mm-wavelength 
interferometry to identify the correct multi-wavelength counterparts,
as has been done with the 
Plateau de Bure Interferometer (e.g. Downes et al.\ 1999) and the
Submillimeter Array (Younger et al.\ 2007). Given
the small field of view available with submm/mm-wavelength
interferometers (typically $<$1$'$ diameter at wavelengths
shorter than $\sim$1.2~mm) and present-day sensitivities, a more practical approach is to use
radio interferometry, where the deepest 1.4~GHz Very Large Array
(VLA\footnote{The NRAO is operated by Associated Universities Inc., under a cooperative
 agreement with the National Science Foundation.}) imaging 
(rms~$\sim$~5--10~$\mu$Jy/beam) is generally found to identify radio counterparts for 
 $\sim$60--80\% of the bright SMGs (e.g. Ivison et al.\ 2005). Many groups have been successful in using
 such deep radio interferometry (usually at 1.4~GHz) to localize the
 counterpart positions with subarcsecond 
 accuracy (e.g. Ivison et al.\ 1998, 2000, 2002, 2007; Smail et al.\
2000; Webb et al.\ 2003; Clements et al.\ 2004; 
 Borys et al.\ 2004; Dannerbauer et al.\ 2004), which relies on the
 locally observed FIR-to-radio luminosity 
correlation (Condon 1992) holding at high-redshifts, 
 which appears to be the case for FIR-selected
 star-forming galaxies out to redshifts, $z \la 1$ (Appleton et al.\ 2004; 
 Kovacs et al.\ 2006; Ibar et al.\ 2008).
Those submm/mm source candidates without radio counterparts are thought to be 
 either spurious, to contain very cold dust, or to lie at redshifts $> 3$.

The steepness of the submm/mm source counts means that wide-area
surveys ($>$1000 sq. arcminutes) are needed to identify members of the   
 extremely bright SMG population ($S_{\rm 1.2mm} \ga 10$~mJy).
One of the largest such surveys to date has been
 conducted with the 1.2~mm MAx-Planck Millimeter Bolometer (MAMBO) camera 
 (Kreysa et al.\ 1998, 2002) toward the field of the A2125 cluster at $z \sim 0.25$ 
(15$^h$~41$^m$, +66$^o$~18$^m$), covering 1600~sq. arcmins.
%(Voss et al.\ 2006). 
 The region centered on the A2125 cluster has been the target of a
 multi-wavelength observing campaign from  
 X-ray to radio wavelengths (Wang et al.\ 2004; Owen et al.\ 2005a,
 2005b, 2006; Voss et al.\ 2006). The main focus of the
 present study is on the \textbf{brightest,} mm-selected high-redshift
 starburst galaxies
 and AGN \textbf{over the entire 1600~sq. arcmins surveyed by MAMBO}. 
Although the large-scale overdense structure occupies most
 of the field, gravitational lensing of the background sources is 
 likely to be unimportant, except near the cluster core, which only
 occupies $\sim$1 sq. arcmin. of the entire 1.2~mm MAMBO map. 
 MAMBO identified four unusually bright mm sources with 1.2~mm flux
densities in the range, $10 - 100$~mJy, of which three were deduced to 
 be radio-loud quasars based on their significant X-ray
 luminosities and variability at  mm-wavelengths. They also exhibit 
 flat-spectrum mm-to-radio flux densities implying a non-thermal origin
 for the mm emission (Voss et al.\ 2006).  
  The fourth object, MMJ1541+6630 (hereafter J154137+6630.5), is
  proposed to be a starburst-dominated galaxy, possibly with a weak
  obscured AGN. Its AGN nature has been concluded from a tentative identification of an X-ray 
 counterpart in the low angular resolution \textit{ROSAT} PSPC map
 (Voss et al.\ 2006).

In this paper we report on 1.4~GHz radio and \textit{Chandra} X-ray observations (where available) of the 
most robust 1.2~mm sources identified in the A2125 field. In
\S\ref{sec:obs} we present the mm, radio and X-ray data used in our
study, while in \S\ref{sec:analyse} we estimate the radio-to-mm photometric redshifts and 
 the X-ray luminosities of our 1.2~mm sample. Finally, in section \S\ref{sec:discuss} we discuss the 
 implications of our results. 
Throughout this work we adopt a cosmological model with 
$H_0 = 71$~km~s$^{-1}$~Mpc$^{-1}$, $\Omega_{\rm m} = 0.27$ and $\Omega_{\Lambda} = 0.73$
(Spergel et al.\ 2007).

\section{Observations}
\label{sec:obs}

\subsection{Previous 1.2~mm MAMBO and 1.4~GHz VLA imaging}

The MAMBO 37 and 117-element bolometer arrays (Kreysa et al.\ 1998,
2002) on the IRAM 30~m telescope
 were used to map the A2125 cluster field at 1.2~mm between 1999 and 2004. The total on-source 
 integration time was 220 hours and the preliminary findings of this
 survey are presented by Bertoldi et al.\ (2000), Voss et al.\ (2006),
 and Sawitzki et al.\ (2009). The 1-$\sigma$ depth of the final 1600~arcmin$^2$ map ranges from 
 $\sim$0.6~mJy/beam ($\sim$11$''$)  
 at the center, to $\sim$3~mJy/beam around the edges. The data analysis and source extraction methods are
 described in detail by Sawitzki et al.\ (2009). Here, we consider the radio and X-ray properties of the 
29 highest signal-to-noise 1.2~mm source candidates ($\ga$4-$\sigma$) identified in the 
MAMBO map (Table~\ref{tab:a2125sources}). \textbf{Relative to the
  cluster core, the nearest 1.2~mm source in our sample is separated by
  3.4~arcmins, while the majority have separations greater
  than 10~arcmins. Using a lense model for the A2218 galaxy cluster, 
 Knudsen et al.\ (2006) show that gravitational lensing is important
 for submm sources in the redshift range, $z = 1 - 4$, if they are 
 within the central few square arcmins of the field centered on the
 cluster. As A2125 has a similar mass and redshift to that of A2218 
 (Wang, Owen \& Ludlow 2004), we do not believe lensing will bias
 our analysis of the 1.2~mm sources in the A2125 field.}

The A2125 field has also been imaged at 1.4~GHz with the VLA in all four configurations to a 
 maximum depth of $\sim$6~$\mu$Jy/beam, while the synthesized beam size is $1\farcs60 \times1\farcs52$ 
(position angle 87.9$^{o}$) over an area of $\sim$40$'$~$\times$~40$'$. 
 These observations, and the subsequent data analysis procedure is described by 
 Owen et al.\ (2005a).

\subsection{\textit{Chandra} X-ray data}

The analysis of an 82~ks \textit{Chandra} ACIS-I image centered on the A2125 cluster is presented by 
Wang et al.\ (2004). These data successfully detect one of the three quasars, J153959+6605.8 (MMJ1540+6605), 
 at X-ray wavelengths. Within the central 17$' \times 17'$ region of the field, these data detect 99 
 discrete X-ray sources, of which 10 are found to be radio-bright and
 physically associated with the A2125 complex.

\begin{figure}
\centering
\includegraphics[width=5.5in]{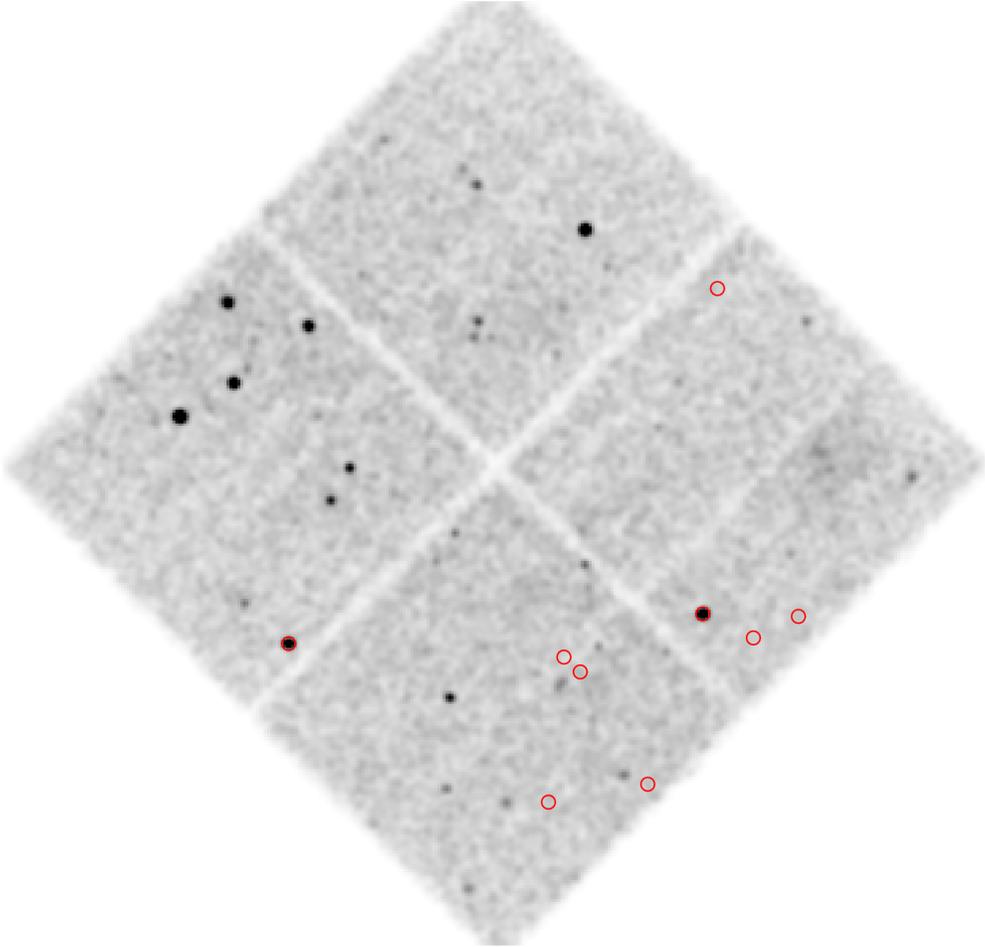}
\caption{
31~ks \textit{Chandra} exposure of the field offset from the center of the A2125
cluster. Circles of radius 10$''$ indicate the positions of radio
counterparts to MAMBO sources in the field. 
The center of the image is 15$^h$42$^m$31$^s$ +66$^d$26$^m$10$^s$(J2000).}
\label{fig:a2125chandra}
\end{figure}

 To study the X-ray properties of the high-redshift starburst galaxy,  J154137+6630.5, as well 
as the other two mm-selected quasars in the A2125 field, a 31~ks \textit{Chandra} ACIS-I exposure 
 was obtained on April 2, 2006, for a field centered $\sim$17$'$ to the North-East of
 the deeper \textit{Chandra} exposure. 
These new data were further analyzed using the \textit{Chandra} Interactive
Analysis of Observations (CIAO\footnotetext{See
  http://cxc.harvard.edu/ciao for a description of CIAO.}) Version 4.0 tools after initial
processing by the \textit{Chandra} X-ray Center (CXC). The smoothed map is shown in 
Figure~\ref{fig:a2125chandra}.  Sources were identified using \textit{\rm CELLDETECT}.

\section{Analysis}
\label{sec:analyse}

\subsection{Identification of 1.4~GHz counterparts to MAMBO sources}

 We outline the criteria adopted for selecting the most probable radio 
counterparts to the 1.2~mm MAMBO source candidates in the A2125 field. We begin by searching 
a 5$''$ radius around the positional centroid of each of the 23 MAMBO source candidates, 
proposing that any 1.4~GHz radio source detected at $\ge$4-$\sigma$ (where $\sigma$ is measured
 locally) within that
 area is a potential counterpart. The search radius is chosen
   to be half of the FWHM of the MAMBO beam at 1.2~mm, a value which has
 been demonstrated to be an appropriate search radius for previous studies of
 850~$\mu$m SCUBA sources (see Ivison et al.\ 2007). 
 We then follow the 
 method of Downes et al.\ (1986) to quantify the statistical
 probability, $P_c$, that the radio counterpart could 
 appear by chance given the local density of radio sources (see also
 Ivison et al.\ 2002, 2007). A radio counterpart with $P_c < 0.05$ is considered to be a 
secure association. Using this approach, 
 we find potential radio counterparts for 24 out of the 29 MAMBO
 source candidates (Table~\ref{tab:a2125sources}). 
Multiple radio counterparts are identified for three of our sources; 
 J154004+6610.3, J153957+6613.6 and J154050+6608.6. 
 Of the 5 mm sources without radio identifications, 3 
(J153935+6629.7, J154249+6625.4 and J154239+6615.5) are found in
 regions of high noise, while two are found close to the edge of the
 MAMBO map. If these mm sources were spurious, then our detection rate
 of robust 1.2~mm sources would be 92\%, consistent with 
 the highest fraction of radio sources idenitified in
  other submm/mm-wavelength surveys (e.g. Ivison et al.\ 2005).

\begin{figure}
\centering
\includegraphics[width=5.0in]{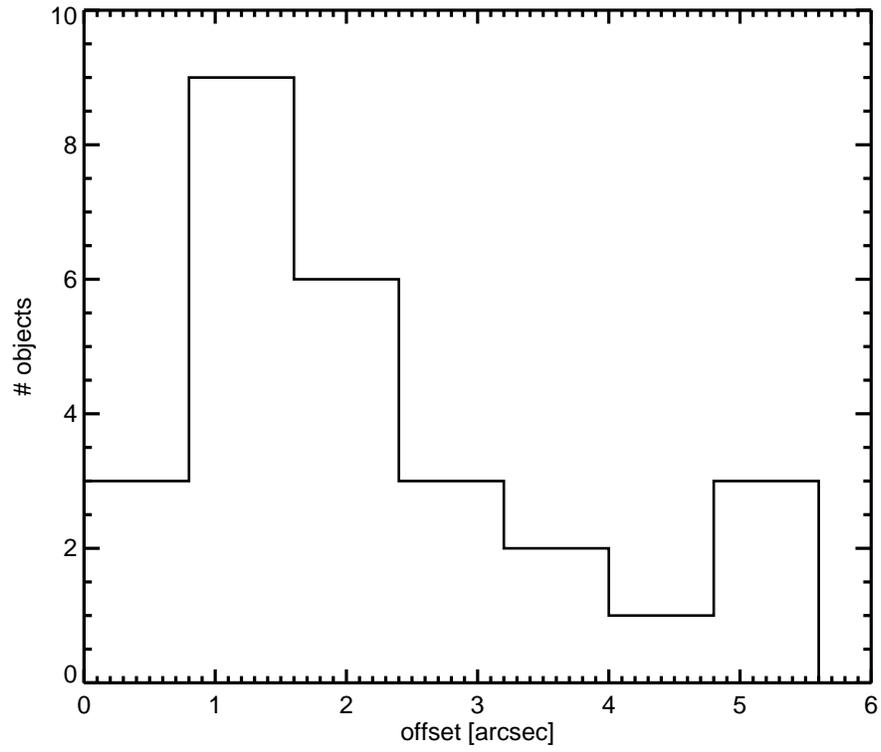}
\caption{Offsets between the mm and 1.4~GHz positions of the proposed radio counterparts to the 
1200~$\mu$m source candidates in the A2125 field.}
\label{fig:offsets}
\end{figure}

The median offset between the positional centroids of our mm and radio source candidates is 1.9$''$ 
(Figure~\ref{fig:offsets}), a value in agreement with that found by Ivison et al.\ (2005) for
 their sample of robust SMGs (selected by comparing SCUBA and MAMBO source candidates)
with secure radio counterparts in the Lockman Hole field.

\subsection{Photometric redshift estimates}

The radio-to-mm  wavelength spectral indices of high-redshift SMGs can serve as a crude 
 indicator of their redshifts. We apply the technique of Carilli \& Yun (1999, 2000) to derive the 
 most probable redshift for each of the 1.4~GHz detected, non-quasar source candidates  in our sample
 (Table~\ref{tab:a2125sources}).  The median  photometric redshift of these
 radio-detected SMGs (excluding the three quasars) is $z \sim 2.2$
 (first and third quartile values of 1.7 and 3.0), 
 which is consistent with the median redshift measured spectroscopically for bright, 
 radio-detected SMGs
  by Chapman et al. (2003, 2005), and that derived for 850~$\mu$m
 selected sources in the  SCUBA HAlf Degree
Extragalactic Survey (SHADES;  Mortier et al.\ 2005; Coppin et al.\ 2006) adopting a 
 more sophisticated photometric redshift technique (Aretxaga et al.\ 2007).

\subsection{X-ray luminosities}

The total 0.5--8.0~keV X-ray luminosity of a galaxy is defined as: 
\begin{equation}
\label{eq:lx}
L_{0.5-8keV} = 4 \pi d_l^2 f_{0.5 - 8keV} (1 + z)^{\Gamma - 2}~erg~s^{-1},  
\end{equation} where $f_{0.5 - 8keV}$ is the 0.5-8~keV flux density (ergs~cm$^{-2}$~s$^{-1}$)
and $d_l$ is the luminosity distance (cm), while here we assume a photon index, 
$\Gamma=1.8$, typical of unabsorbed AGNs (Tozzi et al. 2006).
This X-ray luminosity can be used as a discriminant between starburst and AGN activity, as  
even for moderate redshifts, few starburst galaxies exhibit X-ray luminosities in excess of, 
$L_{0.5-8keV} > $10$^{42}$~ergs~s$^{-1}$ (e.g. Moran, Lehnert, \& Helfand
 1999; Zezas, Alonso-Herrero, \& Ward 2001; Alexander et al. 2002).  
Two of the three high-redshift, mm-selected quasars originally identified by Voss et al.\ (2006), are  
confirmed to be X-ray luminous in our new \textit{Chandra} data
(J154321+6621.9 and J154141+6622.6; Table~\ref{tab:a2125xray}). 
 The third quasar was previously detected in 
 the 82~ks \textit{Chandra} exposure presented by Wang et al.\ (2004).

The fourth bright mm source, J154137+6630.5 (MMJ1541+6630), is not 
 detected in our high resolution \textit{Chandra} data above a 0.5--8.0~keV flux of 
  7$\times$10$^{-16}$~erg/cm$^2$/s. 
Only one other MAMBO source candidate,  
 A2125\_1200.22, is detected with the new X-ray data. The majority of the 1.2~mm source candidates 
 are not detected in these X-ray data, which is consistent with the interpretation that most of their
 FIR luminosity is powered by star-formation, rather than AGN activity. 
 Table~\ref{tab:a2125xray} gives the X-ray properties of the four mm
sources detected in our \textit{Chandra} image. It is also possible that 
 a non-detection of these SMGs in the X-ray bands is indicative of Compton-thick 
 AGN activity.

For systems heavily obscured by dust and gas, strong X-ray emission is hardened due to 
reprocessing of the soft X-ray photons, which are absorbed and re-radiated in the (rest-frame) FIR. 
Quantitatively, the hardness ratio is
calculated from the counts in the hard 2.0--8.0~keV band ($H$) to those in the
soft 0.5--2.0~keV band ($S$), and is defined here as: $(H-S)/(H+S)$. Table~\ref{tab:a2125xray} gives the 
 hardness ratios of the three mm-selected quasars in the A2125 field.

\section{Discussion}
\label{sec:discuss}

\subsection{J154137+6630.5: an obscured starburst galaxy}

 Radio-to-mm wavelength photometric redshift estimates for
 J154137+6630.5 predict, $z = 3.93^{+1.11}_{-0.80}$,
  meaning that it is likely the highest 
 redshift source in our sample, and potentially the most FIR luminous
 SMG discovered in any blank-field submm/mm-wavelength survey 
 ($\sim$5$\times$10$^{13}$~$L_{\odot}$). 
 An X-ray counterpart to J154137+6630.5 is not detected in our new \textit{Chandra} data,  implying,
 $L_{0.5-8.0keV} < (6.2--19.3)\times 10^{43}$~erg~s$^{-1}$. This limit is consistent with the 
X-ray luminosities of the non-AGN SMGs detected in the 
2~Ms \textit{Chandra} survey of the \textit{Hubble Deep Field}-North
  (HDF-N; Alexander et al.\ 2003), as well as low luminosity AGN.  
We therefore suggest that most of the far-infrared luminosity in J154137+6630.5 is 
likely powered by star-formation rather than AGN activity, however it is still possible that
 a Compton-thick AGN is present. If an AGN were present in J154137+6630.5, then correction for 
 the AGN contribution to the radio flux density would result in a higher photometric redshift 
estimate for this object.
 If all of the FIR luminosity indicated by the 1.2~mm flux density were due to star-formation,
  the implied star-formation rate would be a tremendous, $\sim$18,000~$M_{\odot}$~yr$^{-1}$.

\subsection{Radio identification rate and implications for the
  redshift distribution of bright SMGs}

The fraction of robust radio counterparts identified 
 for the 1.2~mm source candidates (24/29) is similar to that typically found in blank-field 
submm/mm-wavelength surveys of comparable depths ($\le$85\%;
 e.g. Ivison et al.\ 2005). This would suggest 
that the redshifts for the majority of our radio-detected sample are at $z \la 3.5$, given the typical
 radio flux densities of these counterparts.

 As mentioned previously, 3 of the 5 mm source candidates without 
 radio counterparts are found in noisier regions of the MAMBO map. 
 Based on previous surveys, it has been argued that some of those 
submm/mm source candidates without radio identifications are likely to be spurious. 
 This charge motivated Ivison et al.\ (2005) to compile a 
sample of robust SMGs using a technique which combines
 source candidates identified in independent surveys of the Lockman Hole (at both 850~$\mu$m
 and 1.2~mm), leading to a reduced likelihood of spurious
 sources in their final catalogue ($\sim$10\%). They then
 find that 80\% of the sources in their catalogue have robust radio 
 counterparts, so that only $\sim$10\% of their candidates are likely to lie at 
redshifts, $> 3$, in agreement with the results presented here.   
Along this same vein, for the SHADES survey, four independent analyses of the same
 dataset are performed in order to minimize systematic uncertainties introduced through the methods  
employed when calibrating  low signal-to-noise bolometer data, effects which may have led to spurious
source candidates in previous SCUBA surveys. The 
fraction of 850~$\mu$m SHADES source candidates with robust 
 radio identifications is $\sim$66\% (Ivison et al.\ 2007), while the depth of the radio imaging is comparable
 between the SHADES and A2125 surveys. 
The typical 1.2~mm rms of the A2125 map is 
  1.0~mJy, while that of the 850~$\mu$m SHADES imaging is 2.2~mJy. Assuming a greybody spectrum 
 with emissivity index, $\beta = 1.5$, and dust temperature, $T_d = 35$~K, typical of the bright SMG 
 population (Kovacs et al.\ 2006; Laurent et al.\ 2006; Coppin et al.\ 2008), we can determine whether the 
 relative depths of the two surveys should be sensitive to the same population of objects over all redshifts. 
For this assumed spectral energy distribution (SED), the MAMBO survey would be sensitive to a typical 
SHADES source
 out to redshifts, $z \la 3.6$, beyond which the 850~$\mu$m continuum emission is expected to be too faint 
 at the limit of the SHADES survey. Conversely, the 1.2~mm emission should become brighter than the 
 MAMBO detection threshold for SHADES sources at higher redshifts. Therefore, given our assumed SED, the
1.2~mm A2125 survey should be more sensitive to higher redshift SMGs, and we would expect a 
\textit{lower} radio detection rate than in the 850~$\mu$m SHADES survey if all source candidates are 
 indeed real. Given the small areas covered by these two surveys, it may be that we
 are effected by cosmic variance in this sample comparison, and that a larger fraction
 of 1.2~mm A2125 sources are at lower redshifts than the SHADES
 sources. If gravitational lensing by the foreground cluster has
 amplified the flux densities of the 1.2~mm source candidates, then
 this comparison would be biased, though we do not believe this
 to be the case.  
 It is also possible that a higher fraction of the
 SHADES sources may be spurious than has been estimated. If this is the case,
 then it is unlikely that more than $\sim$10--15\% of bright SMGs exist
 beyond redshifts, $z > 3.5$, supporting the previous claim by Ivison et al.\ (2005).

Given the high density of radio sources associated with the  
A2125 cluster (Owen et al.\ 1999), it is possible that some of
our proposed radio counterparts to the 
 mm source candidates are in fact associated with foreground cluster members, rather than 
intrinsic to the higher redshift host galaxies responsible for the mm emission. In order 
 to determine if this is indeed the case, we have compared our list of radio identifications 
 with that of the radio positions of cluster members. Only 
 one of our possible radio counterparts to a 1.2~mm MAMBO source candidate is spatially coincident
 with a cluster member, namely the second radio counterpart to J1540041+6610.3 associated with the cluster 
member, 24027 at $z = 0.2425$ (Owen et al.\ 2005b). However, this source is unlikely to be the correct 
 counterpart given the far better positional agreement between the mm
 source and the primary radio counterpart. 
 This primary radio counterpart to J1540041+6610.3 is detected in the \textit{Chandra} X-ray image, and the 
implications of this are discussed below. 
 Independant of which radio/X-ray counterpart gives rise to the mm
 emission in J1540041+6610.3, we can conclude that 
robust radio identifications are been 
 obtained for at least 83\% of our 1.2~mm source candidates in the
 A2125 field.

\subsection{Luminous mm-selected quasars}

The higher angular resolution \textit{Chandra} imaging presented here
 and in Wang et al.\ (2004), confirms the significant X-ray 
 fluxes of the three quasars identified by Voss et al.\ (2006). Two of these 
  (J154321+6621.9 and  J154141+6622.6) have spectroscopic redshifts measured 
 from optical emission lines (Miller et al.\ 2004; Voss et al.\ 2006).
 In the case of the third quasar,  J153959+6605.8, Voss et al.\ (2006) argue that
 the optical spectral energy distribution is consistent with $z =
 0.5$, which we adopt in our luminosity calculations
 (Table~\ref{tab:a2125xray}). All of the quasars have intrinsic X-ray
 luminosities in the range, 
10$^{44}$--10$^{45}$~ergs~s$^{-1}$, consistent with the interpretation
that AGN are powering their tremendous X-ray luminosities. 
If the X-ray emission were due to star-formation
activity, then these luminosities would imply rates in excess of 
10,000~$M_{\odot}$~yr$^{-1}$  (Grimm et al.\ 2003).

For an intrinsic power-law photon index $\Gamma = 1.8$, the hardness ratios of
J154321+6621.9 and  J154141+6622.6 are suggestive of moderate column
densities of obscuring gas ($N_{HI}<$5$\times$10$^{22}$~cm$^{-2}$) at their measured redshifts. 
In the case of J153959+6605.8, its hardness ratio and assumed redshift
($z \sim 0.5$) is consistent with that of an unobscured AGN with an HI
column density, $N_{HI}\sim$10$^{21}$~cm$^{-2}$.

\subsection{J154004+6610.3: a FIR luminous X-ray source}

\begin{figure}
\centering
\includegraphics[width=5.in]{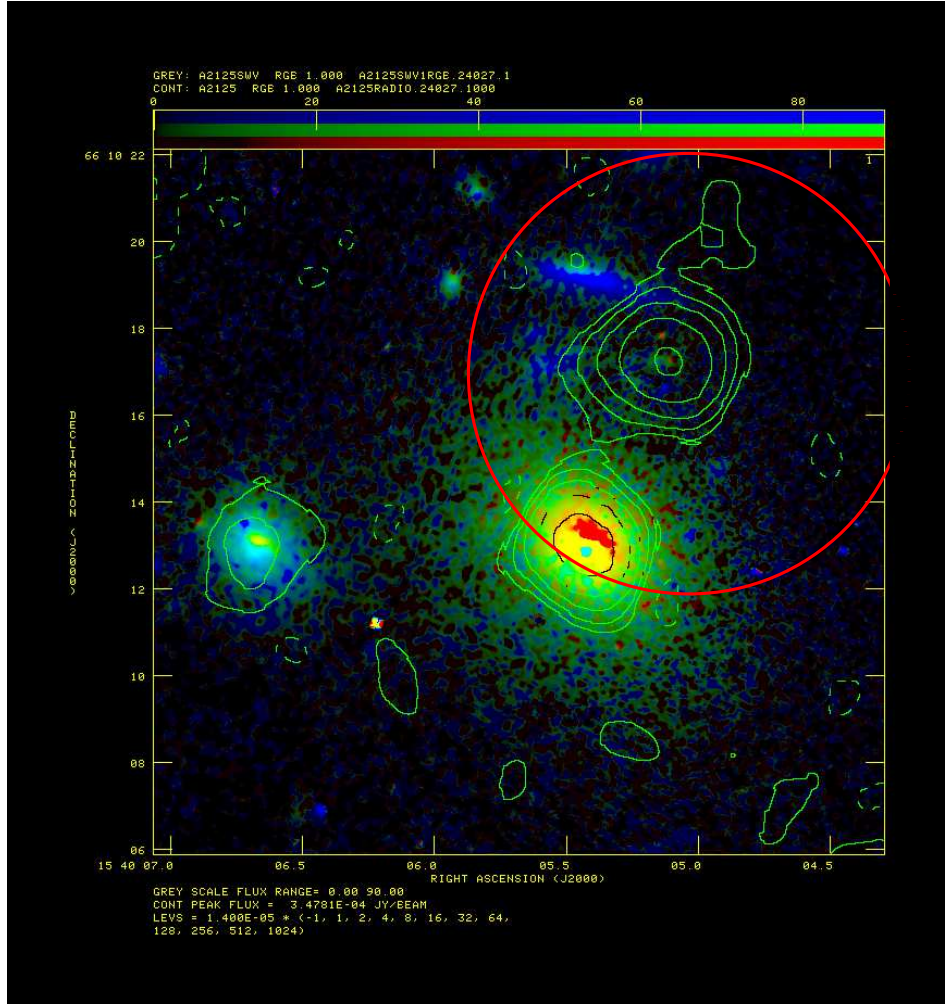}
\caption{
1.4~GHz Radio contours overlaid on an HST `truecolor' image composed of
\textit{V} to \textit{I}-band imaging of the J154004+6610.3
pair. Radio contours intervals are (-1, 1, 2, 4, 8, 16, 32)$\times$14$\mu$Jy/beam. 
The large circle shows the positional centroid and 5$''$ radius region centered on the
1200~$\mu$m MAMBO source candidate. }
\label{fig:a2125p22}
\end{figure}

In addition to the three quasars, the only other mm source in the A2125 field with a proposed X-ray 
 counterpart is J154004+6610.3. The 1.2~mm source candidate is positionally coincident with two radio 
counterparts (Figure~\ref{fig:a2125p22}), and each of these is associated with a \textit{Chandra} 
X-ray source (Wang et al.\ 2004). While one counterpart is a cluster
 member (Owen et al.\ 2005b), both counterparts are spatially coincident with the concentration of 
low surface brightness X-ray emission (LSBXE) to the South-West of the A2125 cluster (Wang et al.\ 2004). 
The \textit{HST} \textit{V}-band image of 
this pair (Figure~\ref{fig:a2125p22}) reveals that the Southern X-ray component is associated with
 a bright elliptical galaxy (Wang et al.\ 2004) at a redshift z=0.2425.
The optically faint component to the North is therefore most likely
  responsible for both the 1.2~mm emission
 and the second X-ray/radio component, which may be lensed by
 the foreground galaxy. The correct counterpart to the mm source
 candidate could be confirmed by submm/mm-wavelength interferometry. 
 In the likely scenario whereby all of the mm emission arises
 from this second source, then the photometric redshift estimate
 predicts z=1.20$^{+0.32}_{-0.25}$. Assuming the range of
 redshifts consistent with the photometric redshift estimate, the
 derived X-ray luminosity of the Northern counterpart
 (Table~\ref{tab:a2125xray}) is in the range, 
$L_{0.5-8keV} \sim (1.4 - 4.5)\times
 10^{43}$~ergs~s$^{-1}$, which would suggest 
 an AGN-dominated system, if the source is unlensed. 
 This interpretation is consistent with
 the point-like nature of the optical counterpart. 
 The X-ray luminosity of the Southern counterpart ($L_{0.5-8.0keV} = 6.3\times
 10^{41}$~erg~s$^{-1}$) is consistent with star-formation, but more
 likely arises from a weak AGN, as only a small fraction of
 such early-type galaxies generally contain detectable quantities of
 molecular gas (e.g. Combes et al.\ 2007), which is required to fuel a starburst.

\section{Summary}

We have presented
 the radio and X-ray properties of the 24 radio-detected 1.2~mm source
 candidates of the 29 detected in the A2125 field mapped by MAMBO. Our main
 findings can be summarized as follows:

\begin{enumerate}

\item The bright thermal source discovered in the 1.2~mm MAMBO
  map is not detected in these new X-ray data, despite a previous
  claim of an identification in the \textit{ROSAT} PSPC
  catalogue. The non-AGN nature of the FIR emission is therefore best
  explained by starburst activity, with an expected star-formation
  rate, $>$10,000~$M_{\odot}$~yr$^{-1}$. Based solely on its 1.2~mm 
  flux density, J154137+6630.5 is therefore
  likely to be the most luminous SMG identified in any blank-field submm/mm
  survey thus far.

\item Our 1.4~GHz VLA imaging reveals an 83\% detection rate for the
 mm source candidates, consistent with previous studies of bright SMGs
 in blank-field surveys of comparable sensitivity, suggesting that most of these
 objects lie at redshifts, $z \la 3.5$. The median mm-to-radio photometric 
redshift of this radio-detected sample is $z \sim 2.2$, consistent 
with that of the median redshift of SMGs with optical spectroscopic redshifts 
(Chapman et al.\ 2003, 2005).

\item The three mm-selected quasars discovered by Voss et al.\ (2006)
  are detected in these  \textit{Chandra} data, 
 while their X-ray luminosities are consistent with AGN activity.

\item X-ray emission is found to be associated with the most likely
  radio counterpart to one of the mm source candidates with a photometric 
 redshift, $z \sim 1$.

\end{enumerate}

Given the  faintness of the optical and mid-infrared counterparts to
J154137+6630.5  (not detected above an 
 \textit{R}-band magnitude of 27.2), the most probable route to
 obtaining a redshift for 
this object is through a broadband mm/cm-wavelength spectroscopic search 
 for  redshifted molecular CO line emission (e.g. Wagg et. al.\
 2007).  Planned Submillimeter Array observations will soon reveal whether the
 single-dish mm emission from this object is composed of multiple
 components, providing an alternative explanation for its large
 apparent FIR luminosity.

With the recent advent of \textit{Spitzer}, it has become possible to
identify counterparts to SMGs in the mid-infrared (Egami et
 al.\ 2004; Frayer et al.\ 2004; Ashby et al.\ 2006; Pope et al.\ 2006; 
 Ivison et al.\ 2007), which may also provide an alternative diagnostic
 of AGN activity. 
We will present an analysis of the mid-infrared properties of 
1.2~mm source candidates in the A2125 field in a forthcoming article.
Within the next two years, the EVLA will come online and  provide an
order of magnitude increase in 1.4~GHz continuum sensitivity. At such
depths it will likely be possible to detect radio counterparts for
 essentially all of the bright SMG population ($\ga$5~mJy at 850~$\mu$m),
depending on the nature of the relatively unstudied high-redshift `tail' of 
 this subsample (e.g. Wang et al.\ 2007). 
However, the steepness of the radio counts at these fainter
  flux density levels means that multiple plausible counterparts are
  likely to exist for a single bright SMG. 
 As such, combined studies using both the EVLA and ALMA will
be necesary to fully understand the submm and radio properties of
this high-redshift galaxy population.

\section{Acknowledgments}

JW thanks Elizabeth Galle for assistance with the \textit{Chandra} data
analysis. JW and CC are grateful for 
support from the Max-Planck Society and the Alexander von Humboldt Foundation.

\begin{landscape}
\begin{center}
\begin{deluxetable}{lcccccccc}
\tablecaption{MAMBO positions and properties of radio-detected 1.2~mm
  star-forming galaxies in the A2125 field. 
\label{tab:a2125sources}}
\tablehead{
\colhead{Source} & 
\colhead{RA}  & 
\colhead{DEC (J2000)} & 
\colhead{S$_{1.2mm}$ [mJy]} & 
\colhead{S$_{1.4GHz}$ [$\mu$Jy]} & 
\colhead{$\triangle$RA [$''$]} &
\colhead{$\triangle$DEC [$''$]} &
\colhead{$P_c$} &
\colhead{z$_{phot}$}$^{a}$ 
}
\startdata
 J153843+6610.2 & 15:38:43.77 & +66:10:12.33 &  8.9$\pm$2.0 &
 96.0$\pm$21.3 & 0.5 & 2.4 & 0.003 & 3.05$^{+1.18}_{-0.78}$ \\
 J153951+6610.3 & 15:39:51.24 & +66:10:15.07 &  4.2$\pm$0.9 &
 39.5$\pm$9.3  & -1.7 & -1.3 & 0.008 & 3.27$^{+1.32}_{-0.89}$  \\
 J153957+6613.6 & 15:39:58.17 & +66:13:34.78 &  4.1$\pm$0.7 &
 309.2$\pm$17.2 & -0.5 & -0.7 & 2.4$\times$10$^{-4}$ & 1.19$^{+0.23}_{-0.20}$ \\ 
               &             &              &              &
 84.2$\pm$9.8  & 2.7 & 4.7 & 0.01 &  \\
 J154004+6610.3 & 15:40:05.11 & +66:10:16.68 &  4.0$\pm$0.9 &
 363.4$\pm$10.2 & 0.5 & -0.4 & 1.5$\times$10$^{-4}$ & 0.78$^{+0.22}_{-0.20}$ \\
               &             &              &              &
 533.8$\pm$10.5 & -1.4 & 3.8 & 0.001 &  \\ 
 J154046+6615.9 & 15:40:47.03 & +66:15:52.64 &  3.1$\pm$0.7 &
 108.2$\pm$7.3 & -1.0 & 0.8 & 0.001 & 1.86$^{+0.43}_{-0.36}$ \\ 
 J154050+6608.6 & 15:40:50.21 & +66:20:15.31 &  3.2$\pm$0.6 &
 99.0$\pm$18.4 & 1.2 & 0.2 & 0.001 & 1.58$^{+0.38}_{-0.32}$ \\
               &             &              &              &
 64.3$\pm$8.4  & 4.8 & 0.2 & 0.017 &  \\
 J154101+6618.2 & 15:41:01.93 & +66:18:13.67 &  3.0$\pm$0.7 &
 45.0$\pm$7.4  & -1.6 & -1.4 & 0.009 & 2.65$^{+0.92}_{-0.63}$  \\
 J154115+6607.8 & 15:41:17.97 & +66:22:32.95 &  4.0$\pm$0.6 &
 145.9$\pm$9.8  & 0.7 & -0.8 & 7.2$\times$10$^{-4}$ & 1.82$^{+0.31}_{-0.29}$ \\
 J154123+6616.6 & 15:41:26.60 & +66:14:36.37 &  4.8$\pm$0.7 &
 83.7$\pm$7.1 & -1.8 & -0.9 & 0.004 & 2.49$^{+0.45}_{-0.36}$  \\
 J154127+6616.3 & 15:41:27.50 & +66:16:14.69 &  6.0$\pm$0.7 &
 64.8$\pm$7.4  & 1.3 & -2.3 & 0.008 & 3.05$^{+0.57}_{-0.43}$  \\
 J154128+6622.0 & 15:41:28.33 & +66:21:59.99 &  2.9$\pm$0.7 &
 1331.2$\pm$12.1 & -2.7 & -2.7 & 7.6$\times$10$^{-4}$ & 0.53$^{+0.19}_{-0.17}$ \\
 J154131+6610.9 & 15:41:31.12 & +66:10:50.25 &  3.7$\pm$0.7 &
 39.7$\pm$7.6 & 1.7 & -0.2 & 0.007 & 1.01$^{+0.82}_{-0.61}$  \\
 J154133+6611.9 & 15:41:33.22 & +66:11:58.26 & 2.9$\pm$0.7 &
 54.1$\pm$7.1  & 0.1 & 0.7 & 9.2$\times$10$^{-4}$ & 2.45$^{+0.74}_{-0.60}$  \\
 J154137+6630.5 & 15:41:37.17 & +66:30:31.94 & 15.8$\pm$1.9 &
 97.9$\pm$18.5 & -0.1 & 0.3 & 1.7$\times$10$^{-4}$ & 3.93$^{+1.11}_{-0.80}$ \\
 J154142+6605.9 & 15:41:42.66 & +66:05:57.26 &  3.9$\pm$0.9 &
 95.0$\pm$10.8 & -1.7 & -1.0 & 0.003 & 2.18$^{+0.64}_{-0.49}$ \\
 J154154+6618.5 & 15:41:54.12 & +66:18:30.22 &  3.8$\pm$0.8 &
 37.9$\pm$8.4  & -1.9 & 0.1 & 0.008 & 3.16$^{+1.21}_{-0.82}$  \\
 J154210+6621.2 & 15:42:10.71 & +66:21:12.54 & 3.5$\pm$0.7 &
 173.1$\pm$10.3 & 0.0 & -1.3 & 9.9$\times$10$^{-4}$ & 1.62$^{+0.36}_{-0.32}$ \\
 J154215+6621.6 & 15:42:14.80 & +66:21:35.35 &  2.8$\pm$0.7 &
 74.0$\pm$17.5 & 0.8 & -0.1 & 0.001 & 2.11$^{+0.82}_{-0.61}$ \\
 J154218+6618.1 & 15:42:18.83 & +66:18:04.32 &  2.8$\pm$0.7 &
 68.4$\pm$9.5  & 2.4 & 0.1 & 0.006 & 1.86$^{+0.43}_{-0.36}$ \\
 J154220+6607.2 & 15:42:20.75 & +66:07:12.94 &  4.6$\pm$0.9 &
 66.0$\pm$12.1 & 3.0 & -2.5 & 0.01 & 2.72$^{+0.85}_{-0.64}$ \\
 J154223+6610.1 & 15:42:23.35 & +66:10:03.25 & 2.7$\pm$0.7 &
 120.1$\pm$10.1 & -5.0 & -2.3 & 0.009 & 1.67$^{+0.45}_{-0.37}$ \\
\enddata
\tablenotetext{a~}{Photometric redshifts are calculated following the method of Carilli \& Yun 
 (1999, 2000). For sources with multiple radio IDs, we estimate the redshift by comparing the total radio flux
  density from both counterparts to the 1.2~mm emission. }
\end{deluxetable}
\end{center}
\end{landscape}

\begin{center}
\begin{deluxetable}{lccccc}
\tablecaption{X-ray properties of selected mm sources in the A2125 field. 
\label{tab:a2125xray}}
\tablehead{
\colhead{Source} & 
\colhead{$S_{0.5-8.0keV}$ [counts/ks]} &
\colhead{Band ratio$^a$} &
\colhead{Flux$^b$} &
\colhead{$L_{0.5-8.0keV}$$^{c}$} &
\colhead{Reference$^d$} 
}
\startdata
 J154321+6621.9 & 4.57$\pm$0.48 & -0.30$\pm$0.07  & 57.1   &  43.3 &  (1) \\ %50.9
 J154141+6622.6 & 5.03$\pm$0.52  & -0.20$\pm$0.04 & 62.8   &  63.1 &  (1) \\
 J153959+6605.8 & 11.72$\pm$0.46 & -0.40$\pm$0.05 & 146.3  &  12.8 &  (2) \\
 J154137+6630.5 &  $<$0.06       &    -           & $<$0.7 &  $<$6.2--19.3 &  (1) \\
 J154004+6610.3 & 0.24$\pm$0.07  &    -           & 3.0    &  1.4--4.5 &  (2) \\
                & 0.31$\pm$0.08  &    -           & 3.6    &  0.1 &  (2) \\
\enddata
\tablenotetext{a~}{Hardness ratio calculated from the counts in the 2.0--8.0~keV ($H$) band and those 
 in the 0.5--2.0~keV ($S$) band, defined as $(H-S)/(H+S)$.}
\tablenotetext{b~}{Full-band flux (in units of 10$^{-15}$~erg/cm$^2$/s) calculated by assuming a photon 
 index $\Gamma$=1.4, and a Galactic HI column density $N_{HI} = 2.75\times10^{20}$~cm$^{-2}$.}
\tablenotetext{c~}{X-ray luminosity in the 0.5--8.0~keV band (in units of 10$^{43}$~erg~s$^{-1}$) calculated 
 from Equation~\ref{eq:lx} and assuming $\Gamma=2.0$ for the thermal
 sources and $\Gamma=1.8$ for the three AGN. For
 A2125\_1200.3 we adopt  z=0.5 as is suggested by the optical 
 photometry (Voss et al.\ 2006), while for the other sources we assume either the measured redshift or that 
 estimated from the mm-to-radio photometric redshift technique.}
\tablenotetext{d~}{(1) This work. (2) Wang et al.\ (2004).}
\end{deluxetable}
\end{center}

\clearpage

\end{document}